\documentclass[showpacs,twocolumn,amsmath,superscriptaddress]{revtex4-2}
\usepackage[dvipdfmx]{graphicx}
\usepackage{color,amsmath,lmodern,bm,array,braket}

\newcommand{\hr}[1]{{\color{black} #1}}

\newcolumntype{C}[1]{>{\centering\arraybackslash}p{#1}}
\newcolumntype{L}[1]{>{\raggedright\arraybackslash}p{#1}}
\newcolumntype{R}[1]{>{\raggedleft\arraybackslash}p{#1}}

\begin{document}

\preprint{paper}

\title{{\it Ab initio} exploration of short-pitch skyrmion materials: Role of orbital frustration}%

\author{Takuya Nomoto}%
\email{nomoto@ap.t.u-tokyo.ac.jp}%
\affiliation{Research Center for Advanced Science and Technology, University of Tokyo, Komaba, Meguro-ku, Tokyo 153-8904, Japan}
\author{Ryotaro Arita}%
\email{arita@riken.jp}
\affiliation{Research Center for Advanced Science and Technology, University of Tokyo, Komaba, Meguro-ku, Tokyo 153-8904, Japan}
\affiliation{RIKEN Center for Emergent Matter Science (CEMS), Wako 351-0198, Japan}
\date{\today}%

\begin{abstract}
In recent years, the skyrmion lattice phase with a short lattice constant has attracted attention due to its high skyrmion density, making it a promising option for achieving high-density storage memory and for observing novel phenomena like the quantized topological Hall effect. Unlike conventional non-centrosymmetric systems where the Dzyaloshinsky-Moriya interaction plays a crucial role, the short pitch skyrmion phase requires a quadratic magnetic interaction $J(\bm q)$ with a peak at finite-$\bm Q$, and weak easy-axis magnetic anisotropy is also critical. Thus, conducting first-principles evaluations is essential for understanding the formation mechanism as well as for promoting the discovery of new skyrmion materials. In this {\it Perspective}, we focus on recent developments of the first-principles evaluations of these properties and apply them to the prototype systems Gd$T_2X_2$ and Eu$T_2X_2$, where $T$ denotes a transition metal and $X$ represents Si or Ge. In particular, based on the spin density functional theory with the Hubbard correction combined with the Liechtenstein method in the Wannier tight-binding model formalism, we first show that the Hubbard $U$ and Hund's coupling is essential to stabilize a skyrmion lattice state by enhancing the easy-axis anisotropy. We then discuss mechanisms of finite-${\bm Q}$ instability and show that competition among Gd-5$d$ orbitals determines whether ferromagnetism or a finite-${\bm Q}$ structure is favored in Gd$T_2$Si$_2$ with $T=$ Fe and Ru. \hr{Our systematic calculations reveal that GdRu$_2$$X_2$, GdOs$_2$$X_2$, and GdRe$_2X_2$ are promising, while GdAg$_2X_2$, GdAu$_2X_2$, and EuAg$_2X_2$ are possible candidates as the skyrmion host materials.} Analysis based on a spin spiral calculation for the candidate materials is also presented.
\end{abstract}

\maketitle

\section{Introduction}
A magnetic skyrmion is a particle-like topological spin texture realized in condensed matter systems. Its \hr{magnetization distribution} is characterized by vorticity, helicity, and a winding number, and the coupling to an itinerant electron provides a unique platform for exploring the physics under the emergent gauge field~\cite{Nagaosa,Mertig2021,Tokura2021}. The skyrmion is also considered a candidate element in future magnetoelectric devices due to its stability against an external perturbation and high controllability by an electric current~\cite{Finocchio_2016, Fert2017,Shijiang, Marrows2021, Wang2022}. Particularly, a racetrack memory~\cite{Parkin}, a magnetic random access memory~\cite{Sampaio2013, Zhang2018}, and an artificial synapse for neuromorphic computing~\cite{Huang2017, Song2020} are promising applications of the topologically stable magnetic object. 

The magnetic skyrmions form a two-dimensional lattice in a ferromagnetic background. While an array of magnetic bubbles or skyrmions has been observed in a magnetic thin film since the 1970s~\cite{Lin1973, Doniach1982}, the first observation in a bulk magnet was accomplished in 2009 by the neutron scattering measurement for MnSi with B20 crystal structure~\cite{Muhlbauerb}. In the chiral magnet, an early theory suggested that a Dzyaloshinsky-Moriya (DM) interaction is the key to realizing the skyrmion lattice phase~\cite{Bogdanov,Robler}. Namely, the competition between ferromagnetic and DM interaction induces a non-collinear spin spiral structure, whose modulation pitch is determined by $A/D$ with $A$ being magnetic stiffness and $D$ being a DM constant. Once the spiral state is stabilized, a triple-$\bm Q$ state, which is identical to the triangular skyrmion lattice, is expected to appear under a magnetic field due to the momentum conservation of the quartic term in the Ginzburg-Landau functional. Today, there are many materials showing the DM-induced skyrmion phase in bulk, thin film, and heterostructures. In these systems, a typical modulation pitch is around 10-100nm, much larger than the lattice constant since $D\ll A$ is satisfied in most cases~\cite{Fert1980,Heide,Freimuth_2014,Kikuchi,Koretsune2018}. 

On the other hand, the magnetic skyrmion observed in the centrosymmetric crystals has recently attracted much attention, where the DM interaction should not play any role. The first observation was reported in Gd$_2$PdSi$_3$ with Gd forming a two-dimensional triangular lattice network~\cite{Saha, Kurumaji}, followed by Gd$_3$Ru$_4$Al$_{12}$ with a Kagome lattice~\cite{Max}, GdRu$_2$Si$_2$~\cite{Khanh}, EuAl$_4$~\cite{Shang2021, Kaneko2021, Z2022, Takagi}, and EuGa$_4$~\cite{Z2022, Zhang2022} with a square lattice network. A remarkable feature of the skyrmion lattice in centrosymmetric systems is its relatively shorter modulation pitch, typically given by a few nanometers. Since this does not originate from the DM interaction, its formation mechanism has been intensively studied recently. In terms of the shortness of the modulation pitch, the skyrmion lattices observed in EuPtSi~\cite{Onuki, Sakakibara} and Mn$_{1-x}$Fe$_x$Ge~\cite{Kanazawa2011, Shibata2013} might be classified into the same category as the centrosymmetric systems even though their crystal structures do not have the space inversion symmetry. Since a short-pitch skyrmion lattice is favorable for realizing high-density memory bits in a storage device, revealing the key quantity to determine the modulation pitch will promote further materials design and engineering of the skyrmion phase. \hr{Furthermore, the high skyrmion density of the short-pitch skyrmion is of great theoretical importance, as it provides a unique platform to observe fascinating phenomena, such as a quantized topological Hall effect~\cite{Hamamoto} and a crossover behavior of the topological Hall effect where both the real-space and momentum-space Berry curvature play a crucial role~\cite{Matsui}.}

To date, there have been a lot of theoretical works on the formation of a short-pitch skyrmion lattice. They include calculations based on a classical spin model, an itinerant electron model, and an effective Ginzburg-Landau functional. These theories enable us to grasp a physical insight into the skyrmion lattice formation and, indeed, the importance of the magnetic frustrations~\cite{Okubo2012, Leonov2015}, compass anisotropy of the \hr{magnetic} interactions~\cite{Chen2016, Chen2018}, higher-order spin interactions~\cite{Heinze2011, Hayami2017}, and Fermi surface nesting~\cite{Ozawa2016, Ozawa2017} have been recognized as an essential property to stabilize the skyrmion lattice phase (for more details, see Ref.~\cite{Hayami2021}). However, these model calculations are usually performed with a given modulation vector and include adjustable parameters. Thus, they have no predictive power for the modulation pitch itself in real materials. On the other hand, in spite of the expensive numerical cost, one can investigate the stability of finite-$\bm Q$ spin spiral structures from first-principles, 
which is helpful to see the origin of the modulation pitch and its material dependence. 
Indeed, the calculations for GdRu$_2$Si$_2$~\cite{Nomoto2020, Juba2022}, Gd$_2$PdSi$_3$~\cite{Nomoto2020, Juba2022}, and MnGe~\cite{Grytsiuk2020, Mendive2021} have successfully unveiled microscopic mechanisms in stabilizing their short-pitch skyrmion phase. Thus, broadening the target compounds along this line is of great importance, helping us understand the material dependence more deeply and promoting the discovery of new skyrmion materials. 

\hr{Based on the theory for short-pitch skrymion phase, a promising system for observing skyrmion phases should possess weak easy-axis anisotropy and spin-spin interactions favoring finite-$\bm Q$ modulations~\cite{Hayami2021}. Since these effects on skyrmion formation are well-established, in this {\it Perspective}, we aim to highlight recent progress in first-principles evaluations of these properties and offer predictive insights to guide experimental efforts in discovering new skyrmion materials. To achieve this,} based on spin density functional theory (SDFT) and SDFT with the Hubbard correction $U$ (SDFT+$U$), we perform systematic first-principles calculations for Gd$T_2X_2$ and Eu$T_2X_2$ with $T$ being a transition metal element and $X$ being Si or Ge. Since the largest number of skyrmion compounds have been found in this ThCr$_2$Si$_2$-type crystal structure, these compounds would be ideal target materials. 
First, we show that the inclusion of $U$ and Hund's coupling $J$ is essential to obtain easy-axis anisotropy, which is necessary for stabilizing a skyrmion lattice than a spin spiral state. Then, we evaluate magnetic interactions $J(\bm q)$ based on the so-called Liechtenstein method with the Wannier tight-binding model formalism. The calculated $J(\bm q)$ is in good agreement with $E(\bm q)$ directly evaluated by the spin spiral calculations within SDFT+$U$, supporting the validity of the present Liechtenstein calculations. Based on the results, we argue that the finite-${\bm Q}$ structure is determined not only by the shape of the Fermi surface but also by the details of the electronic structure. Especially, we show that competition among the Gd-5$d$ orbital contributions determines whether a ferromagnetism or finite-${\bm Q}$ structure is favored in Gd$T_2$Si$_2$ with $T=$ Fe and Ru. According to our systematic calculations, GdRu$_2$$X_2$, GdOs$_2$$X_2$, GdW$_2X_2$, GdRe$_2X_2$, GdAg$_2X_2$, GdAu$_2X_2$, EuCo$_2X_2$, and EuAg$_2X_2$ with $X=$ Si and Ge are possible candidates for showing the skyrmion phase. 
The present study provides a firm ground 
to discover and design the short-pitch skyrmion phase from systematic 
calculations, and the extension to the other crystal structures will be a promising future development.

\section{Methods}
In this section, we summarize the methods used in this {\it Perspective}. First, we explain the evaluation of $J(\bm q)$ based on the Liechtenstein method. Although this method was first formulated in the multiple scattering theory with the Green's function techniques, it has been applied to the SDFT Hamiltonian with various spatially localized bases, including the Wannier functions~\cite{Korotin2015, Nomoto2020_2}. Here, we show the formulation following Ref.~\cite{Nomoto2020_2}. We begin with the following tight-binding Hamiltonian,
\begin{align}
H = \sum_{12}(t_{12}+v_{12})c_1^\dagger c_2,\label{eq:hamiltonian}
\end{align}
where indices 1 and 2 run over all degrees of freedom that specify the Wannier functions, including atomic sites $i$, orbitals $\ell$, and spins $\sigma$. $c_1^\dagger$ ($c_1$) represents an electron creation (annihilation) operator in this basis. $t_{12}$ and $v_{12}$ denote spin-independent and spin-dependent hopping integral matrices, respectively. These parameters are extracted from the SDFT/SDFT+$U$ calculation with ferromagnetic reference states through the Wannier construction process~\cite{Mostofi2008, Marzari2012}. 

Based on the Hamiltonian~\eqref{eq:hamiltonian}, we can evaluate the magnetic interaction $J_{ij}$ in the classical Heisenberg model as follows,
\begin{align}
J_{ij}= -\frac{1}{2}{\rm Tr}_{n\ell\sigma}[G_{ij}(\omega_n)\Sigma^{i}G_{ji}(i\omega_n)\Sigma^{j}]. \label{eq:liecht}
\end{align}
Here, ${\rm Tr}_{n\ell\sigma}=T\sum_{n}{\rm Tr}_{\ell\sigma}$, where ${\rm Tr}_{\ell\sigma}$ denotes the trace over the orbital and spin spaces. $\omega_n=\pi T(2n+1)$ denotes the Matsubara frequency, and $T$ is the temperature. The Green's function $G(\omega_n)$ and the magnetic perturbation matrix at $i$ site $\Sigma^{i}$ are defined by,
\begin{align}
[G(\omega_n)]^{-1}_{12}&=i\omega_n\delta_{12}-t_{12}-v_{12}\\
\Sigma_{12}^i&=\tilde{v}_{i\ell_1,i\ell_2}\sigma^x_{\sigma_1\sigma_2}
\end{align}
where we approximate $\Sigma^i_{12}$ as a local quantity. $\sigma^\mu$ ($\mu=x,y,z$) is the Pauli matrix, and $\tilde{v}_{i_1\ell_1,i_2\ell_2}$ is defined by $v_{i_1\ell_1\sigma_1,i_2\ell_2\sigma_2}=\tilde{v}_{i_1\ell_1,i_2\ell_i}\sigma^z_{\sigma_1\sigma_2}$. Then, $G_{ij}(\omega_n)$ is defined as a sub-matrix of $G_{12}(\omega_n)$ having $(i,j)$ sites component.  The evaluation of the Matsubara summation in Eq.~\eqref{eq:liecht} is performed by the sparse sampling technique with the intermediate representation~\cite{Chikano2019, Wallerberger2020}. More details about the implementation are found in Ref.~\cite{Nomoto2020_2}. Based on $J_{ij}$ evaluated by Eq.~\eqref{eq:liecht}, its Fourier transform $J(\bm q)$ is obtained as follows:
\begin{align}
J(\bm q)=\frac{1}{N}\sum_{ij}e^{-i\bm q\cdot(\bm R_i-\bm R_j)}J_{ij} - J_0,\label{eq:fourier}
\end{align}
where $\bm R_i$ denotes the real space coordinate of the site $i$ including the relative coordinate of the sublattice. Here, we have introduced $J_0$, the on-site magnetic interaction, to guarantee the absence of the self-interaction term in the Heisenberg model. Note that, although $J_{ij}$ evaluated by Eq.~\eqref{eq:liecht} will be finite for non-magnetic sites such as a transition metal $T$ and  $X=$ Si, Ge in Gd$T_2X_2$ and Eu$T_2X_2$, the $i,j$ summation in Eq.~\eqref{eq:fourier} should be taken only for the magnetic site, namely, Gd or Eu site in our cases. Indeed, $J_{ij}$ connecting non-magnetic elements can take a large value when both the density of states (DOS) and exchange splitting near the Fermi level are finite, although the mapping into the classical spin model for these orbitals is totally inadequate. This procedure can remove this artificial contribution in the evaluation of $J(\bm q)$, which is essential for the SDFT+$U$ calculations. 
Based on the obtained $J(\bm q)$, we can see whether the system favors the finite-$\bm Q$ modulation. 

On the other hand, we can evaluate the energy of the spin spiral states with the modulation vector $\bm Q$, denoted by $E(\bm q)$, within SDFT and SDFT+$U$. The calculation is based on the generalized Bloch theorem, in which the energy of the spin spiral state is evaluated with specific twisted boundary conditions for the Bloch spinors~\cite{Sandratskii}. A clear advantage of this method is that the calculated $E(\bm q)$ is exact within the SDFT/SDFT+$U$ level, and no assumption of the mapping to the classical spin model is introduced. Moreover, it is not necessary to expand the unit cell even when calculating the finite-$\bm Q$ structure, which makes the calculation with small $\bm Q$ tractable. However, despite these advantages, it requires a calculation for every $\bm Q$, and thus, is much more expensive than the Liechtenstein method if one wants to see the detailed structures of $E(\bm q)$. Thus, we use these two methods in a complementary manner. 

\section{Calculation details}
For the evaluations of $J(\bm q)$ and $E(\bm q)$, we use Vienna {\it ab initio} simulation package (VASP)~\cite{Kresse1996} for the electronic structure calculations. Here, the exchange-correlation functional proposed by Perdew, Burke, and Ernzerhof~\cite{Perdew1996}, and pseudopotentials with the projector augmented wave (PAW) basis~\cite{Bloechel1994, Kresse1999} are used. The spin orbit interaction is neglected except for the magnetocrystalline anisotropy calculations. 

The structure optimization is performed for all Gd$T_2X_2$ and Eu$T_2X_2$ compounds by the SDFT+$U$ method assuming ferromagnetic states. The convergence criteria of the structure optimization is set to $10^{-5}$ eV and the corresponding electronic self-consistent loop is $10^{-6}$ eV with the accurate precision mode. Here, we use a $16\times16\times16$ Monkhorst-Pack $\bm k$-mesh, and a cutoff of the plane wave basis set is set to be twice the recommended maximum value of the cutoff. For the magnetocrystalline anisotropy calculations, we employ $10^{-8}$~eV as the convergence criteria for electronic calculations and $21\times21\times21$ Monkhorst-Pack $\bm k$-mesh. The cutoff is set to be twice the recommended value. For the energy calculations of the spin spiral states, we employ $10^{-8}$~eV as the convergence criteria for electronic calculations, $16\times16\times16$ Monkhorst-Pack $\bm k$-mesh, and the cutoff is 2.5 times as large as the recommended values. 

For constructing the Wannier tight-binding model, we use 156 Bloch states evaluated on $10\times10\times10$ uniform $\bm k$-grid in the disentanglement procedure. The trial orbitals are set to the Gd/Eu-$d$ and $f$ orbitals, transition metal $T$-$s$, $p$, and $d$ orbitals, and $X$ = Si/Ge-$s$ and $p$ orbitals. The outer window is set as it includes all 156 Bloch states, and the inner window is set as it covers from the bottom of the bands to the bands at 4~eV above the Fermi level. The obtained tight-binding model consists of 38 orbitals per spin. Based on the obtained tight-binding model, $J_{ij}$ and $J(\bm q)$ are evaluated by Eqs.~\eqref{eq:liecht} and ~\eqref{eq:fourier}, respectively. Here, we set the temperature $T=58$~K and employ $48\times48\times48$ uniform $\bm k$-grid.

\section{Results and Discussion}
\subsection{Magnetocrystalline anisotropy}

\begin{figure}[t]
\centering
\includegraphics[width=0.40\textwidth, clip]{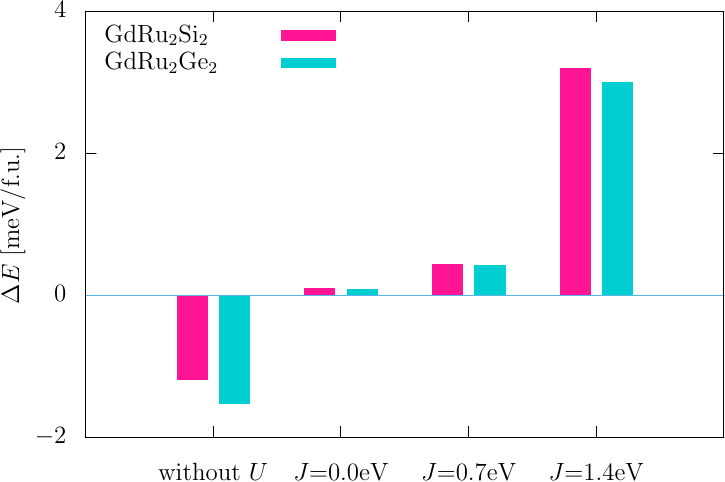}
\caption{
Magnetocrystalline anisotropy of GdRu$_2$Si$_2$ and GdRu$_2$Ge$_2$. The label ``without $U$" stands for the result of the standard SDFT calculation. The others correspond to SDFT+$U$ calculations with $U=6.7$ eV. 
}
\label{fig:mca_gdru2x2}
\end{figure}

According to the theory of skyrmion lattice formation~\cite{Hayami2021}, the system should possess two properties to realize a skyrmion lattice phase in a broad region of parameter space. Namely, it is necessary that the spin interaction is compatible with a finite-$Q$ modulation, and the magnetocrystalline anisotropy is the easy axis type as well as sufficiently weak. \hr{It should be noted that, with the exception of Gd$^{3+}$ and Eu$^{2+}$, rare earth elements possess a finite orbital moment with strong spin-orbit coupling, leading to strong magnetocrystalline anisotropy in most cases. As a result, compounds containing these elements are unlikely to exhibit the weak magnetocrystalline anisotropy required for the formation of skyrmion lattice phases. Therefore, in this {\it Perspective}, we only consider Gd$^{3+}$- and Eu$^{2+}$-based compounds, where the orbital angular momentum of their 4$f$-orbitals are almost quenched in the ionic ground states}. On the other hand, there is no general trend for these elements, whether the anisotropy is the easy axis or easy plane type. Thus, it is crucial to see whether the system shows the correct magnetic anisotropy before discussing the modulation vectors based on the electronic structure. 

\begin{figure*}[!t]
\centering
\includegraphics[width=0.9\textwidth, clip]{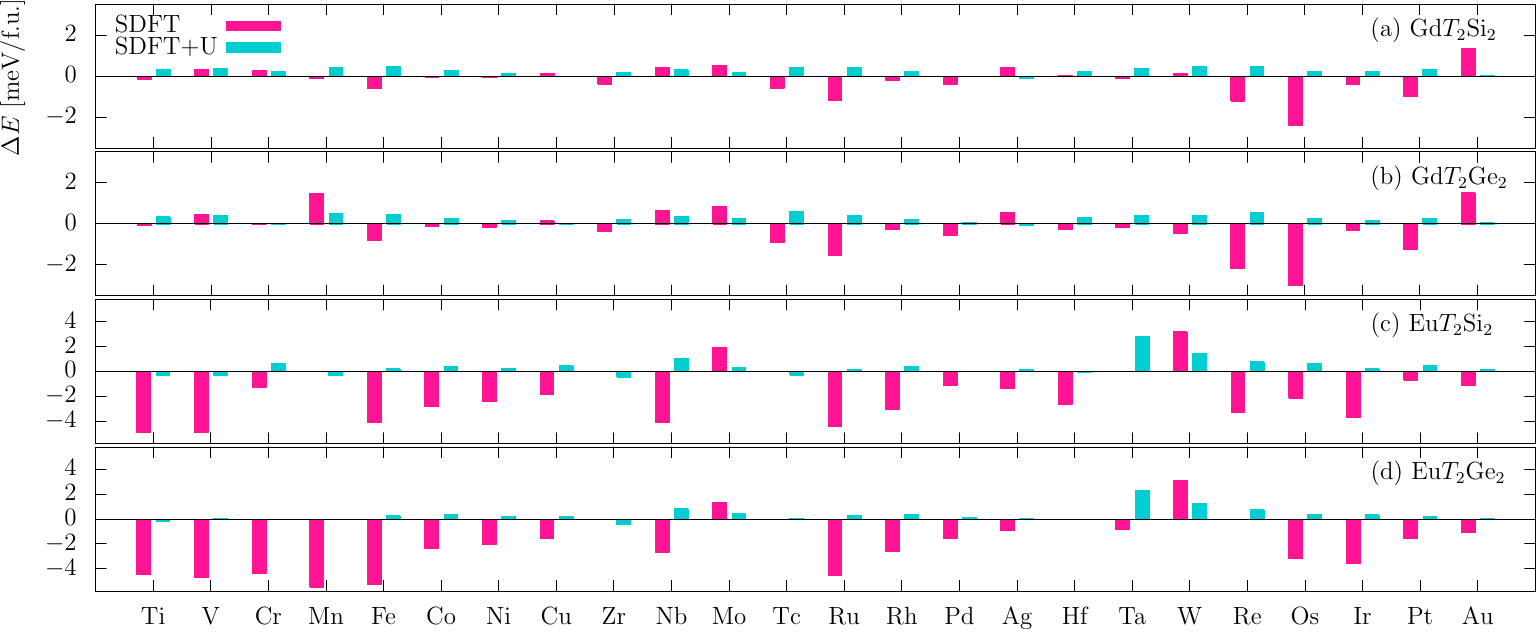}
\caption{
Magnetocrystalline anisotropy of (a) Gd$T_2$Si$_2$, (b) Gd$T_2$Ge$_2$, (c) Eu$T_2$Si$_2$ and (d) Eu$T_2$Ge$_2$. The SDFT+$U$ results stand for the calculations with $U=6.7$ eV and $J=0.7$ eV. Note that the data for EuMn$_2$Si$_2$, EuZr$_2$Si$_2$, EuZr$_2$Ge$_2$, EuTc$_2$Si$_2$, EuTc$_2$Ge$_2$, EuHf$_2$Ge$_2$, EuTa$_2$Si$_2$ and EuRe$_2$Ge$_2$ in SDFT, and GdCr$_2$Ge$_2$ in SDFT+$U$ are not shown. This is because the fully polarized ferromagnetic solutions along the $x$ and/or $z$ axis are unstable for these materials.
}
\label{fig:mca_all}
\end{figure*}

As an example, let us first see in Fig.~\ref{fig:mca_gdru2x2} the calculated magnetocrystalline anisotropy,
\begin{align}
\Delta E:=E(\bm m//\bm x)-E(\bm m//\bm z)
\end{align}
for the skyrmion compound GdRu$_2$Si$_2$ and a reference compound GdRu$_2$Ge$_2$~\cite{Garnier}. We can see that the energy scale of the magnetic anisotropy is quite small, $|\Delta E|<5$ meV per formula unit, as expected. However, the sign of $\Delta E$ is negative in the SDFT results, indicating that a helical magnetic structure is expected to have a lower energy than the skyrmion state. This situation drastically changes by including the Hubbard correction $U$. In the case of Gd$^{3+}$, since $U$ enhances the energy difference between the lowest energy state with $L=0$ and the first excited state with $L\neq 0$, we can expect that the magnetocrystalline anisotropy decreases by increasing the value of $U$, which is consistent with the results in Fig.~\ref{fig:mca_gdru2x2}. Notably, $\Delta E$ becomes small but positive with $U=6.7$ eV even without Hund's coupling correction $J$. In addition, we can also see that the anisotropy becomes larger positive with increasing the strength of $J$, which is compatible with the observed skyrmion lattice phases. This result clearly indicates that the inclusion of $U$ and $J$ is essential to obtain the correct magnetocrystalline anisotropy in the Gd and Eu-based skyrmion materials.

Figures~\ref{fig:mca_all}(a-d) summarize the results of $\Delta E$ in all Gd and Eu-based 122 compounds that we discuss for this {\it Perspective}. We can see that, for most transition metal $T$ except for Ta, the SDFT+$U$ results show the same or lower $|\Delta E|$ than SDFT, which is similar to the cases of GdRu$_2$Si$_2$ and GdRu$_2$Ge$_2$. Although the values of $\Delta E$ highly depend on the host $f$-element and Eu-based compounds tend to have considerable negative $\Delta E$ in SDFT, these trends are strongly suppressed in SDFT+$U$. In SDFT+$U$, $\Delta E$ is no longer sensitive to the transition metal $T$ and the group 14 element $X=$ Si and Ge. Also, the amplitude $|\Delta E|$ is sufficiently small and $|\Delta E| < 3$ meV/f.u. is satisfied in all compounds. It is worth noting that only one (seven) in Gd$T_2X_2$ (Eu$T_2X_2$) among all 48 compounds considered show negative $\Delta E$, implying that the weak easy-axis anisotropy is an intrinsic property inherent in this ThCr$_2$Si$_2$-type crystal structure.

\subsection{Fermi surface and magnetic instability}

Next, let us move on to the modulation vector of Gd$T_2X_2$ and Eu$T_2X_2$. According to the previous discussion, it is essential to include the local Coulomb repulsion $U$ and Hund's coupling $J$ to reproduce the correct magnetocrystalline anisotropy observed in the experiments. Thus, in the present study, we discuss the modulation vector $\bm Q$ based on the electronic structures in the SDFT+$U$ calculations. Apparently, the effect of $U$ and $J$ on $\bm Q$ should not be significant if $\bm Q$ is determined only by the shape of the Fermi surface, which is often supposed in the RKKY scenario for the skyrmion lattice formation. Indeed, since the magnetic moments of Gd$^{3+}$ and Eu$^{2+}$ are fully polarized even in the absence of $U$ and $J$, the Fermi surface is expected to be insensitive to the values of $U$ and $J$. We can directly confirm that magnetic moments are fully polarized in the ferromagnetic states in most compounds except for EuMn$_2$Si$_2$, EuZr$_2$Si$_2$, EuZr$_2$Ge$_2$, EuTc$_2$Si$_2$, EuTc$_2$Ge$_2$, EuHf$_2$Ge$_2$, EuTa$_2$Si$_2$ and EuRe$_2$Ge$_2$. Thus, although the electronic structure above the Fermi energy should be modified to some extent because of the change of the exchange splitting, this will not affect the modulation vector $\bm Q$ in the skyrmion phase. 
 
\begin{figure}[t]
\centering
\includegraphics[width=0.48\textwidth, clip]{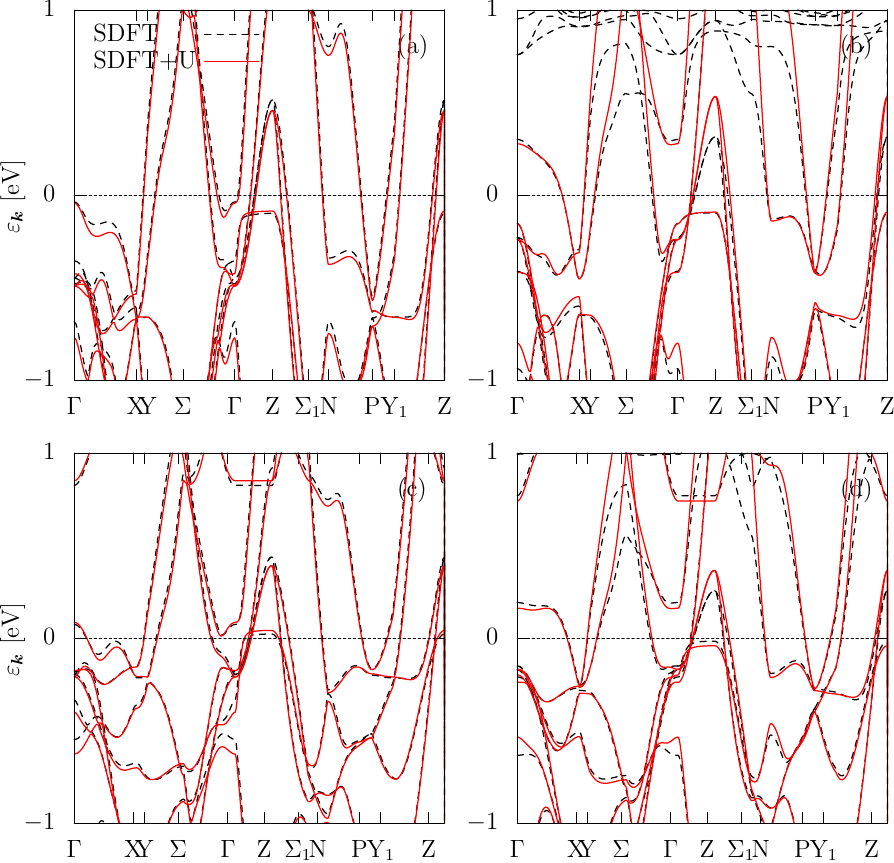}
\caption{
Band structures of (a) majority spin of GdRu$_2$Si$_2$, (b) minority spin of GdRu$_2$Si$_2$, (c) majority spin of GdFe$_2$Si$_2$ and (d) minority spin of GdFe$_2$Si$_2$. Solid lines correspond to SDFT and dashed lines correspond to SDFT+$U$ results with $U=6.7$ eV and $J=0.7$ eV.  
}
\label{fig:bands_feru_fm}
\end{figure}

The insensitivity of the Fermi surface against $U$ and $J$ can be seen directly by the band structure calculations. For example, we show the band structures near the Fermi levels of GdRu$_2$Si$_2$ and GdFe$_2$Si$_2$ in Figs.~\ref{fig:bands_feru_fm}(a-d). Here, the majority and minority bands calculated with SDFT and SDFT+$U$ are plotted, respectively. We can see from the figures that, although they differ in detail especially above the Fermi levels, $U$ and $J$ do not change the Fermi surface, which means that SDFT and SDFT+$U$ results share the same Fermi surface instability. Note that this trend is valid for all compounds when SDFT correctly captures the ionic state of Gd$^{3+}$ and Eu$^{2+}$.

In the continuum model, it is well-known that the Lindhard function describes the Fermi surface instability. In a real material with many bands, the nesting function, 
\begin{align}
\chi_{nn'}({\bm q})=-\frac{1}{N}\sum_{\bm k}\frac{f(\varepsilon_{n{\bm k}+{\bm q}})-f(\varepsilon_{n'{\bm k}})}{\varepsilon_{n{\bm k}+{\bm q}}-\varepsilon_{n'{\bm k}}}, \label{eq:nesting}
\end{align}
plays the same role. Here, $n$ and $\bm k$ denote the band index and crystal momentum, respectively. $\varepsilon_{n{\bm k}}$ is the eigenvalues of the Bloch states having $n$ and $\bm k$, and $f(\varepsilon)$ is the Fermi distribution function. The nesting function is sometimes considered to describe not only the Fermi surface instability but also the spin instability, such as spin-density-wave and helical $\bm Q$ magnetic structures. However, as is known in the fluctuation theory for multi-orbital systems, the orbital character of the bands is essential to describe the fluctuation. In this case, it is necessary to consider the (zeroth-order) spin correlation function $\chi_{\mu\nu}(\bm q)$ defined as follows,
\begin{align}
\chi_{\mu\nu}({\bm q})&=\frac{1}{N}\sum_{\bm R}\sum_{1234}e^{i\phi_{13}^{\bm q}(\bm R)}s^\mu_{13}s^\nu_{24}\chi_{13,24}(\bm R), \label{eq:spinfluc1}\\
\chi_{13,24}(\bm R)&=-T\sum_{\omega_n}G_{12}(i\omega_n,\bm R)G_{43}(i\omega_n,-\bm R). \label{eq:spinfluc2}
\end{align}
Here, the phase $\phi_{12}^{\bm q}(\bm R)$ is defined by $\phi_{12}^{\bm q}(\bm R) = \bm q\cdot(\bm R_{i_3} - \bm R_{i_1})$, and $\omega_n$ and $\omega_q$ denote the Fermionic and Bosonic Matsubara frequency, respectively. Since Eq.~\eqref{eq:spinfluc2} leads to the nesting function with a modification due to the transformation from local orbital to band basis, the $\bm q$-dependence of $\chi_{\mu\nu}({\bm q})$ is mainly determined by Eq.~\eqref{eq:nesting}. However, it should be noted that the basis transformation gives an additional source of the $\bm q$-dependence, and $\chi_{\mu\nu}({\bm q})$ is sensitive not only to the shape of the Fermi surface but also to its orbital character. Moreover, in the strongly correlated electron systems, Eqs.~\eqref{eq:spinfluc1} and \eqref{eq:spinfluc2} are not sufficient to describe the correct spin fluctuation. In this case, the effect of $U$ must be taken more accurately using a diagrammatic technique, which is practically impossible in most cases. In that sense, the Liechtenstein method presented provides a simple but reliable approximation in taking the correlation effects. Indeed, it is known that this method successfully reproduces the result of $t/U$ expansion in the strong correlation limit and reproduces Eqs.~\eqref{eq:spinfluc1} and \eqref{eq:spinfluc2}  in the weak correlation limit. It is worth noting that in the intermediate and strong $U$ regime, not only the Fermi surface but also the electronic structure far from the Fermi level can affect the spin instability.

\subsection{$J(\bm q)$ in GdRu$_2$Si$_2$ and GdFe$_2$Si$_2$ evaluated by the Liechtenstein method}

\begin{figure}[t]
\centering
\includegraphics[width=0.48\textwidth, clip]{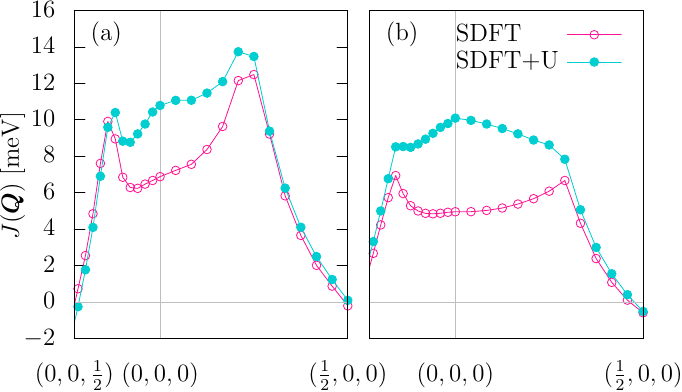}
\caption{
Modulation vector $\bm Q$ dependence of the spin exchange interaction $J(\bm q)$ in (a) GdRu$_2$Si$_2$ and (b) GdFe$_2$Si$_2$ calculated based on the Liechtenstein formula. A pink (cyan) line with open (filled) circle corresponds to the SDFT (SDFT+$U$ with $U=6.7$ eV and $J=0.7$ eV) result.
}
\label{fig:exint_feru}
\end{figure}

Figures~\ref{fig:exint_feru}(a) and (b) show $J(\bm q)$ for  GdRu$_2$Si$_2$ and GdFe$_2$Si$_2$ calculated based on the Liechtenstein formula, Eq.~\eqref{eq:fourier}. The result in Fig.~\ref{fig:exint_feru}(a) shows good agreement with that in Ref.~\cite{Juba2022} but is slightly different from that in
 Ref.~\cite{Nomoto2020} due to an erroneous treatment of the phase factor in Eq.~(\ref{eq:spinfluc1}). According to the results, since the peak position $\bm Q$ in $J(\bm q)$ is the signature of magnetic instability with the modulation vector $\bm Q$, we can say that GdRu$_2$Si$_2$ favors the spin spiral state with $\bm Q\sim (0.25,0,0)$ both in SDFT and SDFT+$U$. On the other hand, in GdFe$_2$Si$_2$,  finite-$\bm Q$ instability is seen only in SDFT while SDFT and SDFT+$U$ give the same Fermi surface (see Figs.~\ref{fig:bands_feru_fm}(c) and (d)). This result indicates that the finite-$\bm Q$ structure is determined not only by the shape of the Fermi surface but also by other factors, such as an orbital character and electronic structure far from the Fermi level. 
 
The ferromagnetic instability observed in the SDFT+$U$ calculation for GdFe$_2$Si$_2$ is also found in the spin spiral calculation (which will be discussed in Section~\ref{subsec:spiral}).
 However, it seems to be inconsistent with the observed spin flop transition in GdFe$_2$Si$_2$~\cite{Mihalik}, which implies
 that we may need to treat the correlation effect more precisely to understand the magnetisn in GdFe$_2$Si$_2$. On the other hand, GdRu$_2$Si$_2$ has a peak at $\bm Q\sim (0.25,0,0)$ both in SDFT+$U$ and SDFT calculation, similar to the previous results based on SDFT (without $U$)~\cite{Nomoto2020, Juba2022}, which is consistent with the experiment.

\begin{figure}[t]
\centering
\includegraphics[width=0.45\textwidth, clip]{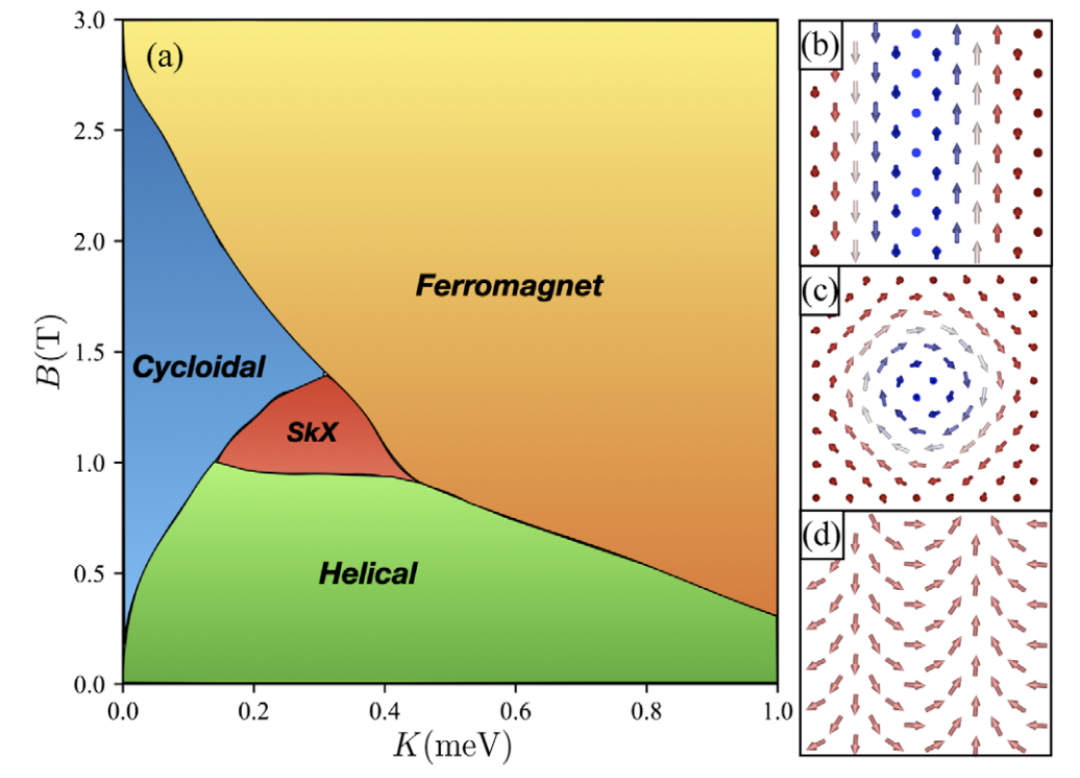}
\caption{
\hr{ (a) A magnetic phase diagram of GdRu$_2$Si$_2$ in the $(K, B)$ parameter space, where $K$ is the magnetocrystalline anisotropy and $B$ is the magnetic field. (b)-(d)
Representative magnetic textures described by the Gd moments. (b) Helical state for $(K,B)=(0.2, 0.6)$, (c) Skyrmion state for
$(K,B)=(0.2, 1.0)$, and (d) cycloidal state for $(K,B)=(0.1, 1.0)$. Reproduced from Ref.~\cite{Juba2022}.}
}
\label{fig:phase}
\end{figure}

\hr{Note that, in principle, it is necessary to perform a spin model simulation to find out a correct magnetic ground state, especially for multiple-$\bm Q$ states like a skyrmion lattice phase. For a spin model of GdRu$_2$Si$_2$, an analysis based on the Landau-Lifshitz-Gilbert (LLG) equation was performed in Ref.~\cite{Juba2022}, using the following model for the internal energy,
\begin{align}
E=-\frac{1}{2}\sum_{ij}J_{ij}{\bm m}_i\cdot{\bm m}_j-K\sum_i(\bm m_i\cdot \bm e_z)^2-\bm B\cdot\sum_i\bm m_i, \label{llg}
\end{align}
where $K$ is a magnetocrystalline anisotropy constant, and $\bm B$ is a external magnetic field applied along the $c$ axis. They solved the LLG equations for Eq.~\eqref{llg} with $J_{ij}$ evaluated from first-principles, which is consistent with Fig.~\ref{fig:exint_feru}(a), and obtained a magnetic phase diagram in terms of $K$ and $B$, as shown in Fig.~\ref{fig:phase}. According to Fig.~\ref{fig:phase}, the skyrmion lattice phase appears when the easy-axis anisotropy $K$ is in the range from 0.2 to 0.4~meV. On the other hand, as discussed in the previous sections, the SDFT+$U$ calculation reproduces the easy-axis anisotropy of GdRu$_2$Si$_2$. The value of $K$ in GdRu$_2$Si$_2$ is $\sim0.43$~meV, which is in good agreement with the spin model simulation Note that the significance of weak easy-axis anisotropy can be straightforwardly understood: if the anisotropy is easy-plane type, a helical single-$\bm Q$ state is favored over multiple-$\bm Q$ states like a skyrmion phase, and if the easy-axis anisotropy is too strong, Ising-type magnetic order is favored over finite-$\bm Q$ structures}. Thus, we conclude that the SDFT+$U$ result in GdRu$_2$Si$_2$ is fully consistent with the observation of the skyrmion lattice phase, and the present method correctly captures the essential physics behind the magnetic structure. 
 

\subsection{Orbital decomposition of $J(\bm q)$ calculated based on the Liechtenstein method}

\begin{figure}[t]
\centering
\includegraphics[width=0.45\textwidth, clip]{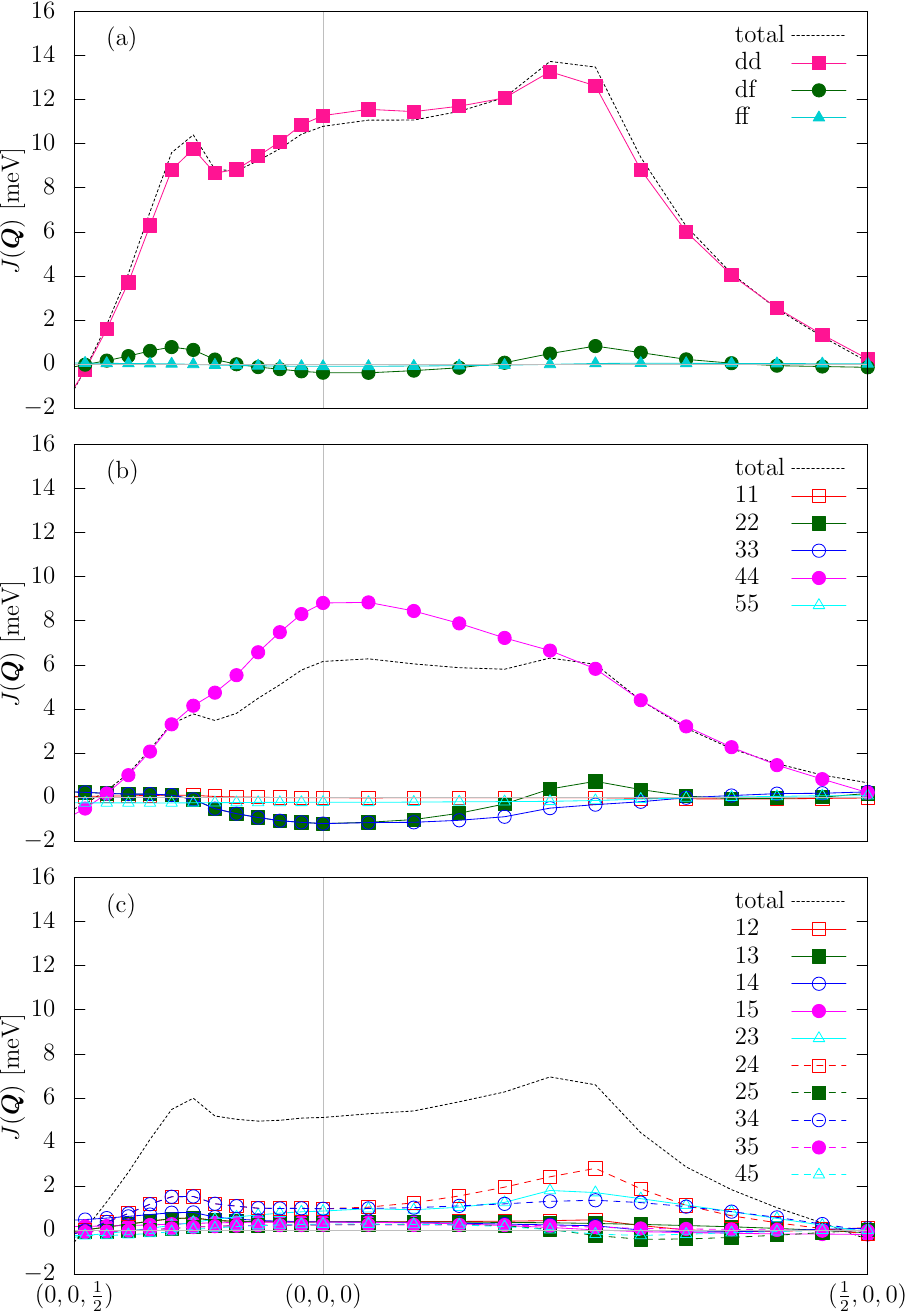}
\caption{
Orbital decomposed $J(\bm q)$ calculated by the Liechtenstein method. (a) The contributions from the Gd-5$d$ and 4$f$ orbitals. (b) The diagonal contribution to in the Gd-5$d$ manifold. Here, the indices 1 to 5 correspond to the $d_{z^2}, d_{xz}, d_{yz}, d_{x^2-y^2}$, and $d_{xy}$ orbitals, respectively. (c) The off-diagonal contributions in the Gd-5$d$ manifold.
}
\label{fig:exint_orb}
\end{figure}

One of the advantages of evaluating $J(\bm q)$ based on the tight-binding formalism is that it is easy to discuss the microscopic origin of the magnetic interaction~\cite{Kvashnin,Nomoto2020}. Although $J(\bm q)$ was decomposed into the Gd-4$f$ and 5$d$ components and a possible frustration mechanism between these two was discussed in Ref.~\cite{Nomoto2020}, one may think that the situation will be different in a real material since their calculations were based on the SDFT electronic structures. Indeed, with the SDFT+$U$ results, we can easily show that the contribution from Gd-4$f$ orbital is strongly suppressed and never affect the modulation vector of the helical states, as shown in Fig.~\ref{fig:exint_orb}(a). Naively, this indicates the absence of conventional RKKY interactions in this compound since it is included in the Gd-4$f$ contribution as discussed in Ref.~\cite{Nomoto2020}. On the other hand, as seen in the previous subsections, the orbital character and the electronic structures not at the Fermi level are essential factors in determining the finite-$\bm Q$ instability. Since the $\bm q$ dependence of $J(\bm q)$ mainly comes from the Gd-5$d$ manifold according to Fig.~\ref{fig:exint_orb}(a), it would be interesting to decompose $J(\bm q)$ into each 5$d$ orbital based on the SDFT+$U$ electronic structures. 

\begin{figure}[t]
\centering
\includegraphics[width=0.42\textwidth, clip]{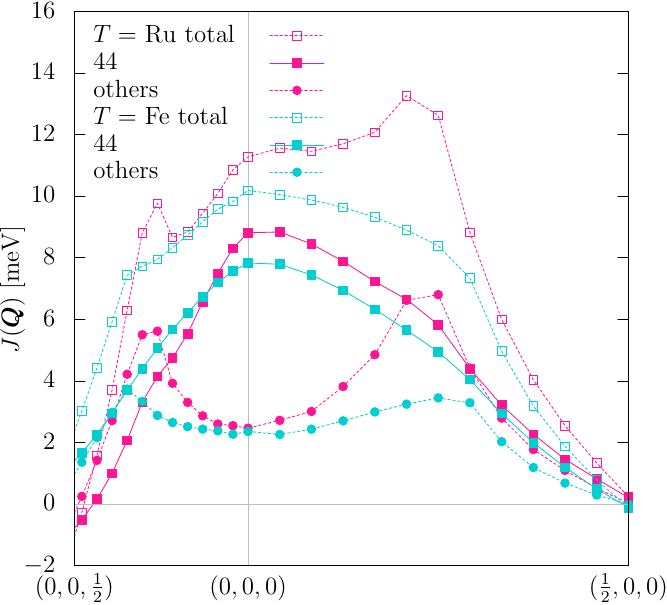}
\caption{
Orbital decomposed $J(\bm q)$ calculated by the Liechtenstein method. The total (dotted line with open square), diagonal $d_{x^2-x^2}$ orbital (solid line with filled square), and the other contribution (dotted like with filled circle) are shown. The pink and cyan lines correspond to the results for GdRu$_2$Si$_2$ and GdFe$_2$Si$_2$, respectively. 
}
\label{fig:exint_orb_v2}
\end{figure}

Figure~\ref{fig:exint_orb}(b) and (c) show the diagonal and off-diagonal contributions to $J(\bm q)$ in the Gd-5$d$ manifold, respectively. Among the five 5$d$ orbitals, the contributions from the $d_{z^2}$ and $d_{xy}$ orbitals are negligibly small, and the $\bm q$ dependence of $J(\bm q)$ originates from the remaining three orbitals. Among the three, we can see that only the diagonal contribution from the $d_{x^2-y^2}$ orbital has a peak structure at the $\Gamma$ point, favoring the ferromagnetic ground state. In contrast, the other diagonal and off-diagonal ones show the peak at $\bm Q\sim(0.25,0,0)$ or almost flat along the $\Gamma$-X line. These results indicate that the orbital frustration exists 
in the 5$d$ orbital manifold, and whether the system favors the ferromagnetic or finite-$\bm Q$ spin spiral state depends on the competition between the contributions from $d_{x^2-y^2}$ and the other orbitals. Notably, this situation is realized not only in GdRu$_2$Si$_2$ but also in GdFe$_2$Si$_2$ as is shown in Fig.~\ref{fig:exint_orb_v2}. The figure shows why these two compounds favor different magnetic structures though they share almost the same Fermi surface. \hr{Namely, the contribution from the $d_{x^2-y^2}$ orbital relative to that from the other $d$ orbitals is larger in GdFe$_2$Si$_2$ than GdRu$_2$Si$_2$. This is mainly due to the suppression of the finite-$\bm Q$ peak in the contribution from the other $d$ orbitals in GdFe$_2$Si$_2$.}

Here, let us look into the detail of the microscopic origin of $J(\bm q)$ based on the following simple model Hamiltonian $H=H_d+H_f+H_{df}$, where,
\begin{align}
H_d&=t_d\sum_{\braket{i,j}\sigma}d^\dagger_{i\sigma} d_{j\sigma} + U_d\sum_i n_{i\uparrow}^dn_{i\downarrow}^d,\label{eq:Hd}\\
H_f&=t_f\sum_{\braket{i,j}\sigma}f^\dagger_{i\sigma} f_{j\sigma} + U_f\sum_i n_{i\uparrow}^fn_{i\downarrow}^f,\label{eq:Hf}\\
H_{df}&=V\sum_{i\sigma}(d^\dagger_{i\sigma} f_{i\sigma} + f^\dagger_{i\sigma} d_{i\sigma})
+J_{cf} \sum_i  {\bm s}^d\cdot{\bm s}^f.\label{eq:Hdf}
\end{align}
Here, $H_d, H_f$ and $H_{df}$ represent terms for the Gd-5$d$ orbitals, 4$f$ orbitals and their hybridization terms, respectively. 
Our calculations for Gd$T_2X_2$ show that, in the presence of $U$, the 5$d$ contribution always overcomes that of 4$f$ since the latter is strongly suppressed. This clearly indicates that both the super-exchange interaction $J_{\rm ex}^f\sim t_f^2/U_f$ and the conventional RKKY interaction $J_{\rm RKKY}\sim (V^2/U_f)^2\chi^{d}$, where $\chi^d$ is the spin susceptibility in the 5$d$ orbitals, are negligibly small in these compounds. Conversely, the strong magnetic interaction from the 5$d$ orbitals can be easily understood since they have large DOS near the Fermi level, and have a finite exchange splitting due to the magnetic ordering although it is not as large as that of the 4$f$ orbitals. Here, we note that there are two possibilities for the origin of the exchange splitting in the 5$d$ states. One is the Coulomb interaction inherent in the 5$d$ orbitals, {\it i.e.}, the second term in eq.~\eqref{eq:Hd}, in which case the spin instability is described purely by the 5$d$ orbitals. The other is the magnetic coupling between the 5$d$ and 4$f$ orbitals, {\it i.e.}, the second term in eq.~\eqref{eq:Hdf}, which one may call "generalized" RKKY interaction since it comes from the coupling between conduction bands and local spins~\cite{Jubacomm}. It should be noted that in SDFT+$U$, the correlation effect is included in the exchange-correlation functional, and the many-body problem turns into an effective one-body calculation. Then, we can not distinguish these two contributions from the resulting exchange splittings. Investigating the microscopic origin of $J(\bm q)$ more precisely requires more elaborated calculations, such as a calculation combined with dynamical mean-field approximation~\cite{Katsnelson}. This direction of development is an interesting and important \hr{future} task in this field.

\begin{figure*}[tb]
\centering
\includegraphics[width=0.73\textwidth, clip]{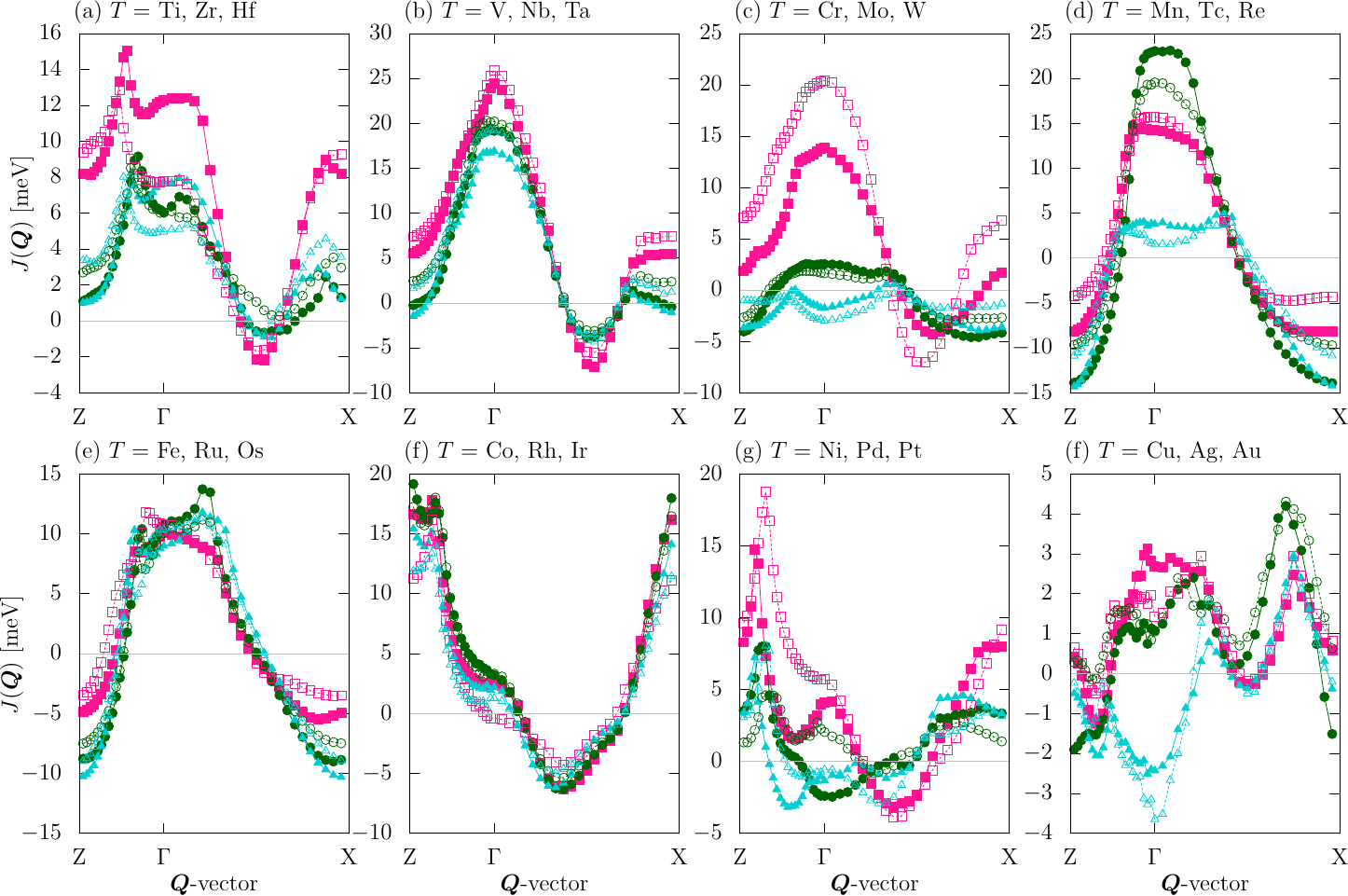}
\caption{
Spin interactions $J(\bm q)$ in Gd$T_2$$X_2$ calculated by the Liechtenstein method based on the SDFT+$U$ electronic structures. The pink, green, and cyan lines with square, circle, and triangle points correspond to 3$d$, 4$d$, and 5$d$ elements as $T$, respectively. The solid (dashed) lines with filled (open) symbols stand for $X=$ Si (Ge). Note that the high symmetry points Z, $\Gamma$, and X correspond to $(0,0,\frac{2\pi}{c}), (0,0,0)$ and $(\frac{2\pi}{a},0,0)$ in the cartesian frame of the reciprocal lattice space, respectively. 
}
\label{fig:exint_gd}
\vspace{5mm}
\includegraphics[width=0.73\textwidth, clip]{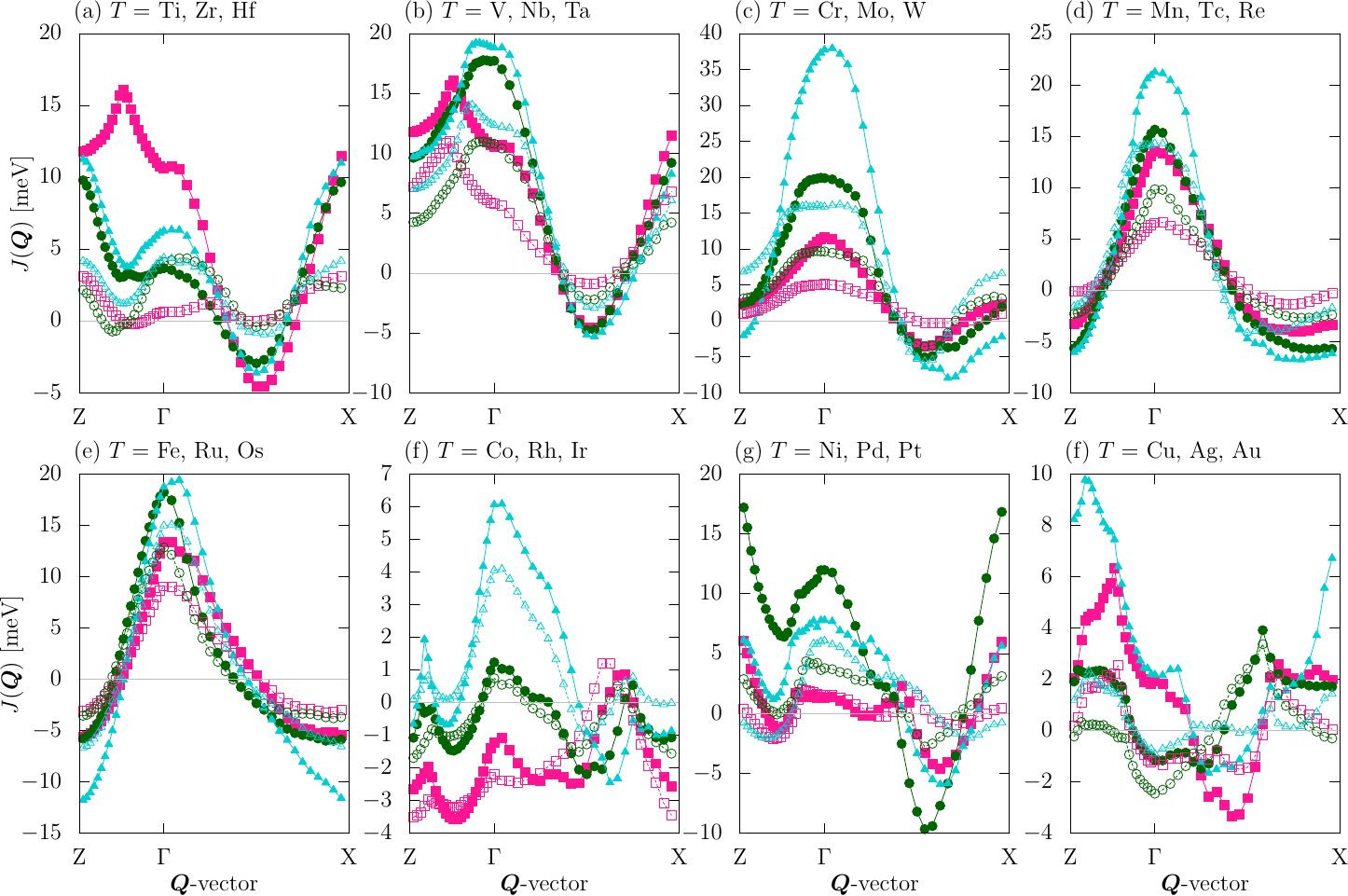}
\caption{
Spin interactions $J(\bm q)$ in Eu$T_2$$X_2$ calculated by the Liechtenstein method based on the SDFT+$U$ electronic structures. Color, line and point types are the same as Fig.~\ref{fig:exint_gd}.
}
\label{fig:exint_eu}
\end{figure*}

\subsection{Systematic evaluation of $J(\bm q)$ based on the Liechtenstein method}

Let us move on to the results of $J(\bm q)$ calculated with SDFT+$U$ for Gd$T_2$$X_2$ and Eu$T_2$$X_2$ in Fig.~\ref{fig:exint_gd} and Fig.~\ref{fig:exint_eu}, respectively. From these figures, 
we see that the general trend of $J(\bm q)$ is not sensitive to the choice of group 14 elements $X=$ Si or Ge. On the other hand, the choices of different $f$-elements and different periods (i.e., 3$d$, 4$d$, or 5$d$) of the transition metals $T$ dramatically change the structure of $J({\bm q})$. For example, for the Gd$T_2X_2$ compounds, the $T$ dependence of group 5 ($T=$ V, Nb, Ta), 8 ($T=$ Fe, Ru, Os), and 9 ($T=$ Co, Rh, Ir) are not so significant. On the other hand, for Eu$T_2X_2$, the $T$ dependence is more noticeable. These dependencies may come from the chemical pressure effects on the lattice structures or the spreads and local energy levels of the $d$-orbitals. 


According to Figs.~\ref{fig:exint_gd} and \ref{fig:exint_eu}, the most promising candidates showing skyrmion lattice phase would be compounds with $T$ being the group 8 transition metal. In particular, GdRu$_2$Si$_2$, GdRu$_2$Ge$_2$, GdOs$_2$Si$_2$, GdOs$_2$Ge$_2$, and EuOs$_2$Si$_2$ have the highest peaks at finite $\bm Q$ along the $\Gamma$-X line. Together with the small easy-axis anisotropy shown in Figs.~\ref{fig:mca_all}(a-d), they should show the skyrmion lattice phase, and indeed, it is already found in GdRu$_2$Si$_2$. Except for these compounds, we can also see that GdW$_2X_2$, GdRe$_2X_2$, GdAg$_2X_2$, GdAu$_2X_2$, EuCo$_2X_2$, and EuAg$_2X_2$ with $X$ = Si and Ge have the highest peaks along the $\Gamma$-X line. However, the peaks in GdW$_2X_2$ and EuCo$_2X_2$ may be too small to stabilize the magnetic order. 

\subsection{Modulation vector based on spin spiral calculations}\label{subsec:spiral}
Up to now, we have discussed the magnetic interaction $J(\bm q)$ and the corresponding modulation vectors $\bm Q$ based on the Liechtenstein method. The calculations are performed for the Wannier tight-binding models derived from first-principles for the ferromagnetic ground states. Thus, in principle, the reliability of the $\bm q$-dependence far from the $\Gamma$ point, as well as the validity of the mapping to the classical spin model cannot be justified. Here, we present spin spiral calculations for the materials expected to be a good candidate for showing the skyrmion phase. Since the calculation is exact within the SDFT+$U$ level, the results can be used in a way complementary to that of the Liechtenstein method.

\begin{figure}[t]
\centering
\includegraphics[width=0.45\textwidth, clip]{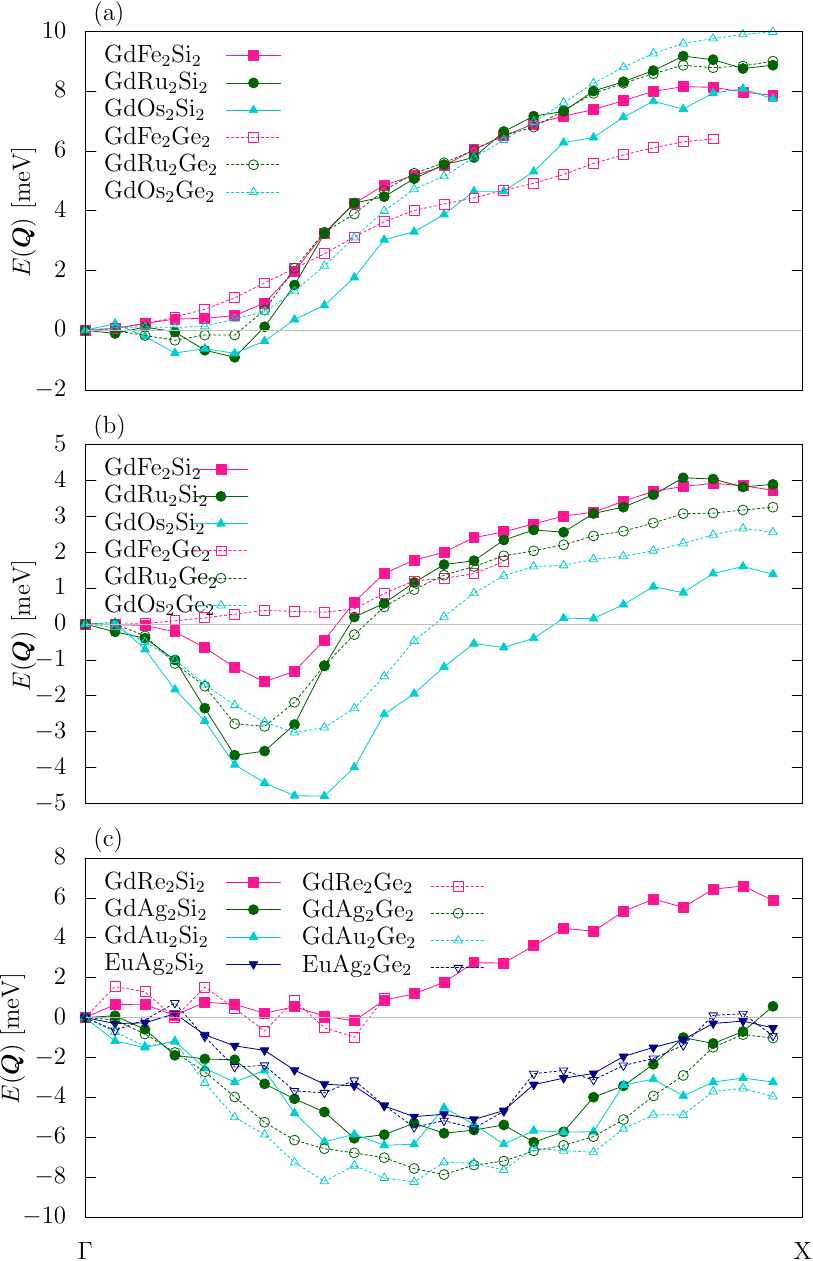}
\caption{
Modulation vector $\bm q$ dependence of the ground state energy $E(\bm q)$ calculated based on the Frozen magnon method. (a) The results based on the SDFT+$U$ with $U=6.7$ eV and $J=0.7$ eV for Gd$T_2X_2$ with $T=$ Fe, Ru, Os and $X=$ Si, Ge. (b) The results based on the SDFT for Gd$T_2X_2$ with $T=$ Fe, Ru, Os and $X=$ Si, Ge. (c) The results based on the SDFT+$U$ with $U=6.7$ eV and $J=0.7$ eV for Gd$T_2X_2$ with $T=$ Re, Ag, Au and $X=$ Si, Ge.
}
\label{fig:frozen}
\end{figure}

Figure~\ref{fig:frozen}(a) shows the $\bm q$-dependence of the total energies for Gd$T_2$$X_2$ with $T=$ Fe, Ru, Os. The calculations are performed by the spin spiral calculations based on SDFT+$U$ with $U=6.7$ and $J=0.7$~eV. Since the minimum position of $E(\bm q)$ represents the most stable modulation vector, we can see that GdRu$_2$Si$_2$, GdRu$_2$Ge$_2$, and  GdOs$_2$Si$_2$ favor the finite-$Q$ state with $\bm Q \sim (0.20,0,0)$-$(0.25,0,0)$. On the other hand, GdFe$_2$Si$_2$ and GdFe$_2$Ge$_2$ have the minimum at the $\Gamma$ point, indicating that the ferromagnetic state is the most stable. These features are in good agreement with the calculations based on the Liechtenstein method. Although the GdOs$_2$Ge$_2$ does not show the finite-$\bm Q$ instability in contrast with the Liechtenstein result, the peak of $J(\bm q)$ in GdOs$_2$Si$_2$ is more fragile than the others as is shown in Fig.~\ref{fig:exint_gd}(e). Thus, the disagreement with the Liechtenstein method is not serious. For the comparison, we also show $E(\bm q)$ of Gd$T_2X_2$ with $T=$ Fe, Ru, Os and $X=$ Si, Ge based on SDFT in Fig.~\ref{fig:frozen}(b). We can see that the finite-$\bm Q$ instability becomes stronger in all cases than in SDFT+$U$. Here, GdFe$_2$Si$_2$ shows the finite-$\bm Q$ instability in SDFT while it does not in SDFT+$U$, which again agrees well with the Liechtenstein calculations. In Fig.~\ref{fig:frozen}(c), we show $E(\bm q)$ for the other candidates for showing the skyrmion phase predicted by the Liechtenstein calculations, namely, Gd$T_2X_2$ with $T=$ Re, Ag, Au and $X=$ Si, Ge, EuAg$_2$Si$_2$ and EuAg$_2$Ge$_2$. We can see that GdRe$_2$Si$_2$, and GdRe$_2$Ge$_2$ show nearly flat $\bm q$ dependence near the $\Gamma$ point, and thus, it is possible that some subtle perturbations select a finite-$\bm Q$ state. On the other hand, GdAg$_2$$X_2$, GdAu$_2$$X_2$, and EuAg$_2$$X_2$ \hr{have a broad peak at $\bm Q\sim(\frac{1}{2},0,0)$ corresponding to the 90-degree rotating structure with the nearest neighboring spins, which seems to be too large to realize a skyrmion phase. However, due to its broad nature, we may still have a chance to achieve a skyrmion phase with smaller $\bm Q$ in experiments by using, for example, chemical substitution and external pressure.} We leave the possible magnetic structures favored in these compounds as a future study. 

\section{Conclusion and outlook}
In this {\it Perspective}, we present systematic first-principles calculations for Gd$T_2X_2$ and Eu$T_2X_2$ with $T$ being a transition metal element and $X$ being Si or Ge. From the magnetocrystalline anisotropy calculations based on SDFT and SDFT+$U$, we show that the inclusion of Coulomb interaction $U$ and Hund's coupling $J$ is essential to obtain the easy-axis anisotropy. Then, based on the SDFT+$U$ electronic structures, we evaluate magnetic interactions $J(\bm q)$ based on the Liechtenstein method and show that the obtained $J(\bm q)$ is in good agreement with $E(\bm q)$ by the spin spiral calculations. Our calculations indicate that the finite-${\bm Q}$ structure is determined not only by the Fermi surface topology but also by the details of the electronic structure, and the competition of each Gd-$5d$ orbital contribution determines whether a ferromagnetic spin configuration or finite-${\bm Q}$ structure is favored in Gd$T_2$Si$_2$ with $T=$ Fe and Ru. According to our calculations, \hr{GdRu$_2X_2$, GdOs$_2X_2$, and GdRe$_2X_2$ are promising candidates, while  GdAg$_2X_2$, GdAu$_2X_2$, and EuAg$_2X_2$ are possible candidates} for showing the skyrmion lattice phase. Since the systematic calculations based on the Liechtenstein method is shown to be helpful in evaluating the finite-$\bm Q$ structure, the extension to other crystal structure would be a possible direction to discover and engineer new skyrmion compounds. \hr{On the other hand, developing new methods to evaluate transport properties in the short pitch skyrmion phase is another crucial direction for its practical applications. Since most of the novel phenomena associated with its high skyrmion density have been studied based on simple models, further development is necessary to achieve a better understanding of material dependence and quantitative evaluations.}

\section{Acknowledgments}
This work was supported by JSPS-KAKENHI (No. 22H00290, 21H04437, 21H04990, and 19H05825) and JST-PRESTO (No. JPMJPR20L7). We also acknowledge the Center for Computational Materials Science, Institute for Materials Research, Tohoku University, for the use of MASAMUNE-IMR (Project No. 202112-SCKXX-0010).

\section{AUTHOR DECLARATIONS}
{\bf Conflict of Interest}

The authors have no conflicts to disclose.

{\bf Author Contributions}

Takuya Nomoto:Formal analysis (lead); Writing - original draft.
Ryotaro Arita: Formal analysis (supporting); Writing - review \& editing.

{\bf DATA AVAILABILITY}

The data that support the findings of this study are available from the corresponding author upon reasonable request.

\bibliography{ref}

\end{document}